\documentclass[a4paper, onecolumn, 11pt,accepted=2023-11-22]{quantumarticle}

\pdfoutput=1

\usepackage[utf8]{inputenc}
\usepackage[english]{babel}
\usepackage[LGR, T1]{fontenc}
\usepackage{amsmath}
\usepackage{float}
\usepackage{graphicx}
\usepackage{amsmath}
\usepackage{amsfonts}
\usepackage{amsthm}
\usepackage{amssymb}
\usepackage{etoolbox}
\usepackage[dvipsnames]{xcolor}
\usepackage{hyperref}
\usepackage{soul}
\usepackage{bbm}
\usepackage{commath}
\usepackage{enumerate}
\usepackage{mathtools}
\usepackage{calrsfs}
\DeclareMathAlphabet{\pazocal}{OMS}{zplm}{m}{n}
\usepackage{upgreek}
\usepackage{calc}
\usepackage{accents}
\usepackage{siunitx}
\usepackage{color}
\usepackage{textcomp}
\usepackage{bm}
\usepackage{dsfont}
\usepackage{relsize}
\usepackage{braket}
\usepackage{enumitem}   
\usepackage{mathrsfs}
\usepackage[symbol]{footmisc}

\usepackage[numbers, sort&compress]{natbib}

\graphicspath{{fig/}} 

\newcommand{\declarebsfgreek}[2]{%
  \protected\csdef{bsf#1}{\mathord{\text{\bsfgreekfont#2}}}%
}
\newcommand{\bsfgreekfont}{\usefont{LGR}{cmss}{bx}{it}}
\declarebsfgreek{alpha}{a}
\declarebsfgreek{beta}{b}
\declarebsfgreek{gamma}{g}
\declarebsfgreek{delta}{d}
\declarebsfgreek{epsilon}{e}
\declarebsfgreek{zeta}{z}
\declarebsfgreek{eta}{h}
\declarebsfgreek{theta}{j}
\declarebsfgreek{iota}{i}
\declarebsfgreek{kappa}{k}
\declarebsfgreek{lambda}{l}
\declarebsfgreek{mu}{m}
\declarebsfgreek{nu}{n}
\declarebsfgreek{xi}{x}
\declarebsfgreek{omicron}{o}
\declarebsfgreek{pi}{p}
\declarebsfgreek{rho}{r}
\declarebsfgreek{sigma}{s}
\declarebsfgreek{tau}{t}
\declarebsfgreek{upsilon}{u}
\declarebsfgreek{phi}{f}
\declarebsfgreek{chi}{q}
\declarebsfgreek{psi}{u}
\declarebsfgreek{omega}{w}

\DeclareMathAlphabet{\pazocal}{OMS}{zplm}{m}{n}

\DeclareMathOperator{\tr}{tr}

\newcommand{\comments}[1]{}
\newcommand{\bea}{\begin{eqnarray}}
\newcommand{\eea}{\end{eqnarray}}

\DeclareMathOperator{\sech}{sech}

\usepackage{environ}
\makeatletter
\NewEnviron{quantifiedequation}[1]{
  \begin{equation}
  \expandafter\make@quantifiedequation\expandafter{\BODY}{#1}
  \end{equation}
}
\newcommand{\make@quantifiedequation}[2]{%
  \m@th 
  \sbox\z@{$\qquad\qquad\displaystyle#2$}
  \sbox\tw@{\let\label\@gobble$\displaystyle#1$}
  \ifdim\dimexpr 1em+\wd\z@+0.5\wd\tw@+2em>0.5\displaywidth
    #2\qquad#1
  \else
    \makebox[0pt][r]{%
      \makebox[\dimexpr0.5\displaywidth-0.5\wd\tw@][l]{\quad\box\z@}%
    }#1
  \fi
}
\makeatother

\renewcommand{\thefootnote}{\arabic{footnote}}

\theoremstyle{plain}

\numberwithin{obs}{section}

\newcommand{\ba}{\begin{align}}
\newcommand{\ea}{\end{align}}

\newcommand{\trB}[1]{\text{tr}_{\scriptscriptstyle B}\left[#1\right]}
\newcommand{\trS}[1]{\text{tr}_{\scriptscriptstyle S}\left[#1\right]}
\newcommand{\pit}{\tilde{\pmb{\pi}}}
\newcommand{\pist}{\tilde{\pmb{\pi}}_{\scriptscriptstyle S}}

\newcommand{\pis}{\pmb{\pi}_{\scriptscriptstyle S}}
\newcommand{\pib}{\pmb{\pi}_{\scriptscriptstyle B}}
\newcommand{\id}{\pmb{\mathbbm{1}}}
\newcommand{\ids}{\pmb{\mathbbm{1}}_{\scriptscriptstyle S}}

\newcommand{\hams}{\pmb H_{\scriptscriptstyle S}}
\newcommand{\logd}{\pmb L_{\scriptscriptstyle S}}
\newcommand{\fish}{\pazocal{F}_{\scriptscriptstyle S}(\beta)}
\newcommand{\fisho}{\pazocal{F}^{(0)}_{\scriptscriptstyle S}(\beta)}
\newcommand{\cfish}{\pazocal{I}_{\pmb{H}_S}(\beta)}
\newcommand{\Xs}{\pmb X_{\scriptscriptstyle S}}
\newcommand{\hamb}{\pmb H_{\scriptscriptstyle B}}
\newcommand{\avs}[1]{\left \langle #1 \right\rangle_{\scriptscriptstyle S}}
\newcommand{\avb}[1]{\left\langle #1 \right\rangle_{\scriptscriptstyle B}}

\newcommand{\parts}{Z_{\scriptscriptstyle S}}
\newcommand{\partb}{Z_{\scriptscriptstyle B}}

\makeatletter
\DeclareFontFamily{OMX}{MnSymbolE}{}
\DeclareSymbolFont{MnLargeSymbols}{OMX}{MnSymbolE}{m}{n}
\SetSymbolFont{MnLargeSymbols}{bold}{OMX}{MnSymbolE}{b}{n}
\DeclareFontShape{OMX}{MnSymbolE}{m}{n}{
	<-6>  MnSymbolE5
	<6-7>  MnSymbolE6
	<7-8>  MnSymbolE7
	<8-9>  MnSymbolE8
	<9-10> MnSymbolE9
	<10-12> MnSymbolE10
	<12->   MnSymbolE12
}{}
\DeclareFontShape{OMX}{MnSymbolE}{b}{n}{
	<-6>  MnSymbolE-Bold5
	<6-7>  MnSymbolE-Bold6
	<7-8>  MnSymbolE-Bold7
	<8-9>  MnSymbolE-Bold8
	<9-10> MnSymbolE-Bold9
	<10-12> MnSymbolE-Bold10
	<12->   MnSymbolE-Bold12
}{}

\newcommand{\ignore}[1]{}
\newcommand{\nobibentry}[1]{{\let\nocite\ignore\bibentry{#1}}}

\newcommand{\bibfnamefont}[1]{#1}
\newcommand{\bibnamefont}[1]{#1}

\newcommand{\op}[1]{\pmb #1}

\newcommand*{\blue}[1]{\textcolor{black}{#1}}


\begin{document}

\title{Energy measurements remain thermometrically optimal beyond weak coupling}

\author{Jonas Glatthard}
\affiliation{Department of Physics and Astronomy, University of Exeter, Exeter EX4 4QL, United Kingdom}
\email{J.Glatthard@exeter.ac.uk}

\author{Karen V. Hovhannisyan}
\affiliation{University of Potsdam, Institute of Physics and Astronomy, Karl-Liebknecht-Str. 24--25, 14476 Potsdam, Germany}

\author{Mart\'{i} Perarnau-Llobet}
\affiliation{D\'epartement de Physique Appliqu\'ee, Universit\'e de Gen\`eve, 1211 Gen\`eve, Switzerland}

\author{Luis A. Correa}
\affiliation{Departamento de Física, Universidad de La Laguna, La Laguna 38203, Spain}
\affiliation{Department of Physics and Astronomy, University of Exeter, Exeter EX4 4QL, United Kingdom}

\author{Harry J. D. Miller}
\affiliation{Department of Physics and Astronomy, The University of Manchester, Manchester M13 9PL, United Kingdom}
\email{harry.miller@manchester.ac.uk}

\begin{abstract}
We develop a general perturbative theory of finite-coupling quantum thermometry up to second order in probe–sample interaction. By assumption, the probe and sample are in thermal equilibrium, so the probe is described by the mean-force Gibbs state. We prove that the ultimate thermometric precision can be achieved—to second order in the coupling—solely by means of local energy measurements on the probe. Hence, seeking to extract temperature information from coherences or devising adaptive schemes confers no practical advantage in this regime. Additionally, we provide a closed-form expression for the quantum Fisher information, which captures the probe’s sensitivity to temperature variations. Finally, we benchmark and illustrate the ease of use of our formulas with two simple examples. \blue{Our formalism makes no assumptions about separation of dynamical timescales or the nature of either the probe or the sample.} Therefore, by providing analytical insight into both the thermal sensitivity and the optimal measurement for achieving it, our results pave the way for quantum thermometry in setups where finite-coupling effects cannot be ignored.
\end{abstract}

\maketitle

\section{Introduction}\label{sec:introduction}

Precise measurements of low temperatures are crucial in experiments dealing with quantum phenomena. Indeed, ultra-low temperatures are routinely achieved in nanoelectronic systems \cite{Sarsby2020, Levitin2022} and cold-atom based platforms \cite{Bloch2005, Chen2020}. Applications range from the study of fundamental problems in condensed matter \cite{Greiner2002, Hasan2010, Nayak2010} to thermalisation in closed quantum systems \cite{langen2013local, langen2015ultracold, bouton2020single}; and from the realisation of thermodynamic cycles \cite{niedenzu2019quantized, barontini2019ultra, bouton2021quantum} to analog quantum simulation in optical lattices \cite{sherson2010single, bloch2012quantum}, and computation on large-scale programmable simulators \cite{ebadi2021quantum, scholl2021quantum}.

On the other hand, the application of estimation-theoretic methods to model thermometric protocols in the ultracold regime has consolidated into the novel field of \textit{quantum thermometry} \cite{DePasquale2018, mehboudi2019review}. Its central goals have been to establish fundamental scaling laws for the signal-to-noise ratio of low-temperature estimates \cite{hovhannisyan2018, potts2019fundamental, Jorgensen2020tight, henao2021thermometric} and to identify design prescriptions that can make a probe more responsive to temperature variations \cite{Correa2015, plodzie2018fermion, mukherjee2019thermometry_control, mitchison2020situ, glatthard2022bending, correa2017thermometry_strongcoupling, seah2019collisional, henao2021thermometric, mok2021optimal, hovhannisyan2021optimalcoarse, Sekatski2022}. A particular focus has been recently placed on sensing applications with atomic impurities \cite{Mehboudi2019, bouton2020single, mitchison2020situ, glatthard2022atom, nettersheim2022sensitivity}, which has helped to close the gap between theory and applications \cite{glatthard2022atom}.

Specifically, the \textit{quantum Fisher information} $\pazocal{F}$~\cite{braunstein1994statistical} has played a central role as a figure of merit for quantum thermometry. It governs the scaling of the best-case signal-to-noise ratio of unbiased temperature estimates, in the asymptotic limit a of large number of measurements both in the frequentist approach, through the Cr\'{a}mer--Rao bound \cite{cramer2016mathematical, rao1945information}, and Bayesian approach, through the van Trees inequality\footnote{The Bayesian framework has been considered within quantum thermometry only recently \cite{johnson2016thermometry,rubio2021global, mehboudi2021fundamental, jorgensen2021bayesian, boeyens2021noninformative, rubio2022scales, alves2021bayesian}, proving particularly useful when estimating temperatures from scarce data in real experimental situations \cite{rubio2021global,glatthard2022atom}.} \cite{vanTrees2004detection, Gill2000}.

\blue{In this paper, we study equilibrium probe-based thermometry \cite{stace2010limits, Correa2015, correa2017thermometry_strongcoupling, hovhannisyan2018, Miller2018, Mehboudi2019, potts2019fundamental, mitchison2020situ}, where the temperature of the sample is inferred by measuring a small system---the probe---which is coupled to it. The total system (sample plus probe) is assumed to be in thermal equilibrium. We compute~$ \pazocal{F}$ in this setting and investigate the most informative measurements for probes that are coupled \textit{non-negligibly} to the sample.}

\blue{We go beyond `weak' probe--sample coupling, by which we mean a `negligibly small' coupling, such that the equilibrium marginal of the probe is simply a Gibbs state with respect to the probe's bare Hamiltonian at the sample's temperature\footnote{\blue{In some cases the term `weak coupling' is used as a synonym of `up to second-order' in the system--bath coupling. Throughout this work, however, we will indicate precisely the order of approximation relevant to each of our results.}}. In this case, temperature may be optimally inferred from energy measurements \cite{Correa2015}, which offers simplicity and universality. This is precisely the steady-state prediction of the common Gorini--Kossakowski--Lindblad--Sudarshan (GKLS) \cite{gorini1976completely, lindblad1976cp, bp} quantum master equation describing the dynamics of the probe. Importantly, however, the GKLS equation is mathematically rigorous only for vanishing probe--sample coupling strength \cite{davies1974markovian}. This means that, strictly speaking, it predicts the correct steady state only in the case of asymptotically vanishing dissipation. Moreover, perturbative quantum master equations may encounter even bigger problems in the context of quantum thermometry. Namely, the assumption that the probe--sample coupling is negligible tends to break down in the low-temperature regime. There, the temperature $T$ sets a small energy scale that can become comparable to the interaction Hamiltonian---no matter how small the latter may be next to the bare Hamiltonians of the sample and the probe \cite{Allahverdyan2002, Allahverdyan2012}. Another important limitation of the weak coupling regime is that there the thermalisation time is generically very long, which may render the equilibrium assumption impractical.}

Here, we refer to the `finite-coupling' regime to denote situations in which the coupling is not negligible. \blue{In this regime, the equilibration time will also be finite, and the equilibrium state of the probe will be described by the marginal of the global probe--sample thermal state}. This is referred-to as \textit{mean-force Gibbs state} and it can be significantly far from being `Gibbsian' with respect to the probe's bare Hamiltonian \cite{Onsager1933, Kirkwood1935, Haake1985, Ferraro2012, thingna2012, Kliesch2014, Hernandez-Santana2015, Miller2018Rev, cresser2021weakandultrastrong, latune2022ultrastrong, timofeev2022hmf, winczewski2021renormalization, Trushechkin2022, Alhambra2022, Becker2022}. While the quantum Fisher information of the probe in this regime has received considerable attention \cite{DePasquale2016local, correa2017thermometry_strongcoupling, DePalma2017, hovhannisyan2018, Miller2018, potts2019fundamental, Mehboudi2019, mitchison2020situ}, not much is known about the optimal thermometric measurement beyond some specific examples \cite{potts2019fundamental, Mehboudi2019, Jorgensen2020tight}. Here, we show that energy measurements \blue{attain optimal \textit{sensitivity} up to the first two leading orders in the coupling strength, even in those cases where the optimal measurement acquires lower-order corrections. It is in this sense that we claim that energy measurements remain thermometrically optimal beyond weak coupling}. This is the first general result about the optimal thermometric measurement beyond the weak-coupling regime and has critical practical importance. Namely, it establishes that non-diagonal measurements, aimed at extracting temperature information encoded in the coherences, do not necessarily confer a \blue{tangible} advantage in the finite-coupling regime, nor does any possible adaptive strategy. Furthermore, our approach provides general formulas to evaluate thermal sensitivity explicitly, to second order in the coupling, circumventing the common pitfalls of the conventional weak coupling assumption.

This paper is structured as follows: In  Sec.~\ref{sec:qt} we introduce the perturbative formalism of finite-coupling quantum thermometry and present our main results. We then apply our results to two standard probe--sample models in Sec.~\ref{sec:examples}; namelly, quantum Brownian motion and the spin--boson model.  In  Sec.~\ref{sec:derivation} we give detailed derivations of the formulas presented. Finally, in Sec.~\ref{sec:conclusions} we summarise and draw our conclusions.

\section{Finite-coupling quantum thermometry}\label{sec:qt}

\subsection{Framework}\label{sec:framework}

Let us consider a probe $ S $ (for `system') coupled to a sample $ B $ (for `bath') with Hamiltonian
\begin{align}\label{eq:hamiltonian}
\op{H}=\hams+\hamb+\op{H}_\text{int}.
\end{align}
The total system is in a Gibbs state at the inverse temperature $\beta$:
\begin{align} \label{eq:globGibbs}
\pit = \frac{e^{-\beta \op{H}}}{\tilde{Z}}, 
\end{align}
with partition function $\tilde{Z}=\tr{e^{-\beta \op{H}}}$, and the goal is to determine the inverse temperature $\beta$ by measuring only the probe $S$. In this paper, we work in the units where $\hbar = k = 1$, the latter being the Boltzmann constant. In all what follows, operators are denoted with boldface symbols. Note that, since only the probe is measured, for our forthcoming results to hold it is only required that the probe's state be given by $\trB{\pit}$---the actual global state may differ from $\pit$ and be microcanonical \cite{Simon, Muller2015, Brandao2015, Gogolin2016, Tasaki2018, Kuwahara2020}, pure \cite{Goldstein2006, Popescu2006, Gogolin2016}, or neither \cite{Brandao2015, Hovhannisyan2022}.

To control the magnitude of the probe--sample interaction, we will henceforth write the interaction term as $\op{H}_\text{int} = \gamma \op{V}$, where the dimensionless parameter $\gamma$ is the coupling strength. The equilibrium state of the probe, from which the inverse temperature $\beta$ is to be inferred,
\begin{align}\label{eq:defmfgs}
\pist=\frac{\trB{e^{-\beta \op{H}}}}{\tilde{Z}}, 
\end{align}
is the so-called mean-force Gibbs state \cite{Miller2018Rev, Trushechkin2022}, and it will only coincide with the bare thermal state of the probe, 
\begin{align}
\pis=\frac{e^{-\beta \hams}}{\parts}, 
\end{align}
in the limit of vanishing coupling $ (\gamma\rightarrow 0) $. Here the partition function is $ \parts=\tr{e^{-\beta \hams}} $.

Estimates for $\beta$, drawn from a measurement record of length $N$ via an unbiased estimator, carry an uncertainty $\delta\beta$ such that the signal-to-noise ratio (SNR) is upper-bounded through the Cram\'er--Rao inequality \cite{cramer2016mathematical, rao1945information, helstrom1969quantum, Holevo1982, braunstein1994statistical} as
\begin{equation}\label{eq:cramer-rao}
	\frac{\beta}{\delta \beta} \leq \sqrt{N \beta^2 \fish}.
\end{equation}
Here, $\fish$ is the quantum Fisher information (QFI), which is given by \cite{helstrom1969quantum, Holevo1982, braunstein1994statistical}
\begin{align}\label{eq:defqfi}
\fish=\trS{\logd^2\pist}, 
\end{align}
where the symmetric logarithmic derivative $\logd$ (SLD) is implicitly defined by
\begin{align}\label{eq:defsld}
    \partial_\beta\, \pist = \frac{1}{2}\lbrace \logd, \pist \rbrace,
\end{align}
with $\{\cdot, \cdot\}$ denoting the anticommutator. Equation \eqref{eq:defsld} is a Lyapunov equation with respect to $\logd$ and, since $\pist$ is a bounded positive operator, it has a unique solution, given by \cite{Bhatia1997a}
\begin{align}\label{eq:lyapunov}
\logd = 2\int^\infty_0 d\lambda \ e^{-\lambda \pist} (\partial_\beta\, \pist) e^{-\lambda \pist}.
\end{align}

The optimal thermometric measurement---namely, the one that minimises $\delta \beta$---is the projection onto the eigenbasis of $\logd$ \cite{helstrom1969quantum, Holevo1982, braunstein1994statistical}. This saturates the Cram\'er--Rao bound \eqref{eq:cramer-rao} in the $N\to \infty$ limit \cite{braunstein1994statistical, Gill2000}. For any other measurement, the best-case scaling of the SNR is given by the `classical' Cram\'er--Rao bound \cite{cramer2016mathematical, rao1945information}. For example, when one projects onto the eigenbasis of $\hams$, then
\begin{equation}
	\frac{\beta}{\delta \beta} \leq \sqrt{N \beta^2 \cfish}, 
\end{equation}
where $ \cfish \leq \fish $ is the (classical) Fisher information associated with this specific measurement. Here, the bound also becomes tight in the $N\to\infty$ limit \cite{Fisher1925}. By definition,
\begin{align}\label{eq:defcfi}
\cfish = \sum\nolimits_n \tilde{p}_n\, (\partial_\beta \ln \tilde p_n)^2, 
\end{align}
where $\tilde p_n :=\bra{\epsilon_n} \pist \ket{\epsilon_n}$ are the populations of the probe's equilibrium state in the eigenbasis $\{\ket{\epsilon_n}\}_n$ of $\hams$. Equivalently, $\cfish$ can be written as
\bea \label{eq:clasSLD}
\cfish = \trS{\mathcal{D}_{\hams}(\logd)^2 \mathcal{D}_{\hams}(\pist)},
\eea
where $\mathcal{D}_{\hams}$ is the dephasing operation in the eigenbasis of $\hams$:
\bea
\mathcal{D}_{\hams}(\op{Q}) := \sum_n \ket{\epsilon_n}\bra{\epsilon_n} \op{Q} \ket{\epsilon_n}\bra{\epsilon_n}.
\eea

The bare energy measurement defined by $\{\ket{\epsilon_n}\}_n$ is known to be optimal in the limit $\gamma \rightarrow 0$, where $\pist \to \pis$ and, therefore, $\logd \to \avs{\hams} - \hams$ \cite{Correa2015}; here and throughout, $\avs{\cdot} := \trS{\pis \, \cdot}$. Furthermore, the bare energy measurement is optimal at any $\gamma$ as long as $[\op{H}_\text{int}, \hams] = 0$, because then $[\pist, \hams] = 0$, and therefore $[\logd, \hams] = 0$, which means that $\{\ket{\epsilon_n}\}_n$ is an eigenbasis for $\logd$. In general, however, $\pist$, and hence $\logd$, will not commute with $\hams$, so the optimal thermometric basis will differ from $\{\ket{\epsilon_n}\}_n$. Nonetheless, as we prove below, the bare energy measurement remains as informative as the truly optimal measurement up to the first two leading orders in the perturbative expansion in the probe--sample coupling strength $\gamma$.

\subsection{Results}\label{sec:results}

\subsubsection{\blue{Optimal sensitivity} of local energy measurements at finite coupling}\label{sec:optimality-general}

Suppose $\pist$ is twice differentiable in $\gamma$. It can then be Taylor-expanded as
\bea\label{eq:expanded-state}
\pist = \pis + \gamma\, \op{p}_1 + \gamma^2 \op{p}_2 + O(\gamma^3),
\eea
where $\op{p}_1$ and $\op{p}_2$ are some operators and $O$ is the `big O' as per the standard asymptotic notation. In the same manner, the SLD $\logd$ of $\pist$ [Eq.~\eqref{eq:lyapunov}] will decompose as
\bea
\logd = -\Delta \hams + \gamma\,\op{l}_1 + \gamma^2 \op{l}_2 + O(\gamma^3),
\eea
where $-\Delta \hams := \avs{\hams} - \hams$ is the SLD of $\pis = \pist\big\vert_{\gamma = 0}$ and $\op{l}_1$, $\op{l}_2$ are some operators related to $\pis$, $\op{p}_1$, and $\op{p}_2$ through Eq.~\eqref{eq:lyapunov}. Plugging these into Eq.~\eqref{eq:defqfi}, we obtain
\begin{multline}\label{eq:fishtayl}
\fish = \, \fisho + \gamma\left[ \tr(\op{p}_1 \Delta \hams^2) - 2 \tr(\op{l}_1 \pis \Delta\hams) \right] \\
+ \gamma^2 \left[ \tr(\op{l}_1^2 \pis) + \tr(\op{p}_2 \Delta\hams^2) - 2 \tr(\op{l}_2 \pis \Delta\hams) - \tr(\{ \op{p}_1, \op{l}_1 \} \Delta\hams) \right] + O(\gamma^3),
\end{multline} 
where $\fisho := \tr(\pis \Delta\hams^2) = \partial_\beta^2 \ln \parts$ is the zeroth-order QFI, i.e., that of $\pis$. Now, observing that 
\begin{align*}
    \tr(\op{p}_1 \Delta \hams^2) &= \tr(\mathcal{D}_{\hams}[\op{p}_1] \Delta \hams^2),
    \\
    \tr(\op{l}_1 \pis \Delta\hams) &= \tr(\mathcal{D}_{\hams}[\op{l}_1] \pis \Delta\hams),
\end{align*}
we immediately conclude from Eq.~\eqref{eq:clasSLD} that, up to the first order in $\gamma$, $\fish$ and $\cfish$ coincide; that is, 
\begin{equation}\label{eq:resultmain0}
    \cfish = \fish + O(\gamma^2).
\end{equation}
However, in general, $\mathcal{D}_{\hams}[\{ \op{p}_1, \op{l}_1 \}] \neq \{ \mathcal{D}_{\hams}[\op{p}_1], \mathcal{D}_{\hams}[\op{l}_1] \}$, meaning that $\fish$ and $\cfish$ do differ in the second order. Therefore, we have shown that the bare energy measurement \blue{reaches optimal thermometric sensitivity} up to the first two leading orders (i.e., zeroth and first order), provided that the $O(\gamma)$ term in Eq.~\eqref{eq:fishtayl} is non-zero. In Appendix~\ref{app:counterexamples} we give explicit examples.

Let us now consider the case where the deviation between the mean-force Gibbs state $\pist$ and the local thermal state $\pis$ appears only at $O(\gamma^2)$. This is relevant from the point of view of applications, since it is the situation most commonly studied in open quantum systems \blue{(see Secs.~\ref{sec:mainformulas} and~\ref{sec:examples})}. Since, in this case $\op{p}_1 = 0$, we have that, due to Eq.~\eqref{eq:lyapunov}, also $\op{l}_1 = 0$. As a result, the $O(\gamma)$ term in Eq.~\eqref{eq:fishtayl} vanishes, so that $\fish$ becomes
\begin{align} \label{zabel}
\fish = \fisho + \gamma^2 \left[ \tr(\op{p}_2 \Delta\hams^2) - 2 \tr(\op{l}_2 \pis \Delta\hams) \right] + O(\gamma^3).   
\end{align}
As above, we have that 
\begin{align*}
    \tr(\op{p}_2 \Delta \hams^2) &= \tr(\mathcal{D}_{\hams}[\op{p}_2] \Delta \hams^2),
    \\
    \tr(\op{l}_2 \pis \Delta\hams) &= \tr(\mathcal{D}_{\hams}[\op{l}_2] \pis \Delta\hams),
\end{align*}
which tells us that, according to Eq.~\eqref{eq:clasSLD}, $\fish$ and $\cfish$ coincide up to the second leading order in $\gamma$. Moreover, as we will show below, if one additionally imposes the generic assumption \eqref{eq:Hint}, the cubic term in Eq.~\eqref{zabel} also vanishes (see Sec.~\ref{sec:derivation}), leaving us with
\begin{align}\label{eq:resultmain} 
\cfish = \fish + O(\gamma^4).
\end{align}

Equations~\eqref{eq:resultmain0} and~\eqref{eq:resultmain} constitute our first main result. Namely, \blue{the thermometric sensitivity of the bare energy measurement is} optimal up to the first two nonzero leading orders in $\gamma$. Importantly, Eq.~\eqref{eq:resultmain} applies to typical open quantum system scenarios, which means that, in such settings, measurement optimisation is only necessary at much stronger couplings\footnote{Note, however, that measurements not commuting with the probe's bare Hamiltonian can be advantageous in \textit{nonequilibrium} thermometry even in the ultraweak coupling limit \cite{Tham2016, Mancino2017}.}.

\blue{Lastly, we emphasise that the deviation of the optimal measurement from the energy measurement can be of lower order in $\gamma$ than the deviation of the optimal sensitivity from the sensitivity provided by the bare-energy measurement. An explicit example of such a situation is constructed in Appendix~\ref{app:counterexamples_B}.}

\subsubsection{Quantum Fisher information in finite-coupling thermometry}
\label{sec:mainformulas}

Having proven that \blue{the sensitivity of} energy measurements remains optimal \blue{beyond vanishing coupling}, we now turn to deriving a closed-form expression for $\fish$ up to the second order in $\gamma$ for the case of $\op{p}_1 = 0$. Here, we outline the derivation of our formulas. For a detailed derivation one may skip directly to Sec.~\ref{sec:derivation}.

We begin by imposing the generic assumption of a factorised probe--sample interaction; namely,
\begin{quantifiedequation}{\mathrm{Assumption \hspace{1mm} I:}}
\op{H}_\text{int} = \gamma\,\op{S}\otimes \op{B},
\label{eq:Hint}
\end{quantifiedequation}
where $\op{S}$ and $\op{B}$ are coupling operators belonging to probe and sample, respectively. Importantly, these are chosen such that $\op{S} \otimes \op{B}$ is of the same dimension as $\hams$ and $\hamb$, to ensure that $\gamma$ is dimensionless.

As we will show below, under Assumption I, the condition $\op{p}_1 = 0$ is equivalent to
\begin{quantifiedequation}{\mathrm{Assumption \hspace{1mm} II:}} \quad \avb{\op{B}} = 0,
\label{eq:avgB}
\end{quantifiedequation}
where $\avb{\op{B}} := \trB{\pib\, \op{B}}$, with $\pib = e^{-\beta\hamb}/Z_B$ being the Gibbs state of the sample with respect to its bare Hamiltonian $\hamb$, and with the same $\beta$ as in Eq.~\eqref{eq:globGibbs}. $Z_B = \text{tr}\, e^{-\beta\hamb}$ is the corresponding partition function\footnote{Note that, whenever $\gamma \neq 0$, the actual reduced state of $B$, $\trS{\pit}$, will \textit{not} coincide with $\pib$. Both $\pib$ and $\pis$ are simply fictitious states that appear in our perturbative expansions.}. Importantly, this assumption is also almost universally adopted in the theory of open quantum systems. There, Assumptions I and II are often disregarded as technical. 

\blue{Note that in many practical situations involving magnetic dipoles in three spatial dimensions \cite{Abragam, Jelezko2006}, the interaction Hamiltonian is of the form $\sum_x \gamma_x \op{S}_x \otimes \op{B}_x$ and cannot be reduced to a simple product form of Eq.~\eqref{eq:Hint}. In such cases, our formulas below can be generalised as long as Assumption II is extended to all $\op{B}_x$'s: $\avb{\op{B}_x} = 0$, $\forall x$. We emphasise that Assumption II is crucial from the perspective of thermometry, and tools additional to the ones employed here are required when it is lifted}\footnote{In open quantum systems theory \cite{bp}, it is often argued that even when Assumption II is not met, it can be easily enforced without loss of generality, by re-defining $\op{H} = \hams'(\beta) + \hamb + \pmb{H}_\text{int}'$, where $ \hams'(\beta) = \hams + \trB{\pib\,\pmb{H}_\text{int}} $ and $ \pmb{H}_\text{int}' = \pmb{H}_\text{int} - \trB{\pib\,\pmb{H}_\text{int}}$. Note, however, that there would be an explicit $\beta$-dependence in the modified probe Hamiltonian, which does crucially change the problem from a thermometric viewpoint.}. \blue{In Appendix~\ref{app:counterexamples} we give explicit examples where Assumption II does not hold.}

Our main tool is the Taylor expansion for the operator exponential function \cite{Araki1973}. Let $\op{Q}$ and $\op{R}$ be a pair of dimensionless linear operators. Then, the function $\mathbb{R}\ni \gamma \mapsto e^{\op{Q}+\xi \op{R}}$ can be cast as \cite{Petz2014a}
\begin{subequations}\label{lem:1}
\begin{align}
e^{\op{Q}+\xi \op{R}}=\sum^\infty_{n=0}\op{Q}_n(1)\xi^n, 
\end{align}
where $\op{Q}_0(s)=e^{s \, \op{Q}}$ and
\begin{align}
\op{Q}_n(s):=\int^s_0 dt_1 \ \int^{t_1}_0 dt_2 \ ......\int^{t_{n-1}}_0 dt_n \ e^{(s-t_1) \, \op{Q}}\, \op{R}\, e^{(t_1-t_2) \,\op{Q}}\op{R}\cdots  \op{R}\, e^{t_n\, \op{Q}}.
\end{align}
\end{subequations}
Applying Eq.~\eqref{lem:1} to $\pist$, we find, after some algebra, that
\begin{align} \label{eq:firstorder}
    \op{p}_1 = - \avb{\op{B}} \pis \int_0^\beta d\beta_1 e^{\beta_1 \hams} [\op{S} - \avs{\op{S}} ] e^{-\beta_1 \hams},
\end{align}
from where it evident that, unless $\op{S} \propto \ids$, $\op{p}_1 = 0$ iff $\avb{\op{B}} = 0$.

Hence, the application of Eq.~\eqref{lem:1} to Eq.~\eqref{eq:defmfgs} up to the second order \blue{in $\xi=\gamma$} results in 
\begin{align}\label{eq:tay1}
\pist=\pis\big(\ids+\gamma^2\Xs\big)+O(\gamma^4), 
\end{align}
where the matrix elements $(\Xs)_{nm} = \bra{\epsilon_n}\Xs\ket{\epsilon_m}$ of the traceless operator $\Xs$ in the eigenbasis of $\hams=\sum\nolimits_n \epsilon_n \ket{\epsilon_n}\bra{\epsilon_n}$ are
\begin{subequations}\label{eq:elements2}
\begin{align}
\begin{split}
(\Xs)_{nn}&= \int^{\beta}_0 du \ (\beta-u)\Phi_{\scriptscriptstyle B}(-iu)\big(\sum\nolimits_k \phi_{nk}(-iu)-\Phi_{\scriptscriptstyle S}(-iu)\big), 
\end{split}\\
\begin{split}
(\Xs)_{n\neq m}&=\frac{1}{\Delta_{mn}}\sum\nolimits_k (\op{S})_{nk}(\op{S})_{km}\, \int_0^\beta du\, \Phi_{\scriptscriptstyle B}(-iu)\, \left( e^{u\Delta_{km}}\, e^{\beta \Delta_{mn}}-e^{u\Delta_{kn}} \right) .
\end{split}
\end{align}
\end{subequations}
We have defined
\begin{align}
    \phi_{nk}(x) = \vert (\op{S})_{nk} \vert^2 e^{-i\, x\, \Delta_{nk}}.
\end{align}
with $\Delta_{nk} = \epsilon_k-\epsilon_n$ and $(\op{S})_{nk}=\bra{\epsilon_n}\op{S}\ket{\epsilon_k} $. We have also introduced here the probe and sample auto-correlation functions, defined as
\begin{subequations}\label{eq:correl}
\begin{align}
    \Phi_{\scriptscriptstyle S}(x)&\coloneqq\avs{e^{ix\hams}\op{S}\, e^{-ix\hams}\op{S}} = \trS{\pis\, e^{ix\hams}\op{S}\, e^{-ix\hams}\op{S}} \\
    \Phi_{\scriptscriptstyle B}(x)&\coloneqq\avb{e^{ix\hamb}\op{B}\, e^{-ix\hamb}\op{B}} = \trB{\pib\, e^{ix\hamb}\op{B}\, e^{-ix\hamb}\op{B}}.
\end{align}
\end{subequations}
From these formulas, we see that the condition for the validity of Eq.~\eqref{eq:tay1}, $\Xs \ll \ids$, is satisfied if $\gamma^2 \langle\op{S}^2\rangle_{\scriptscriptstyle S} \langle \op{B}^2 \rangle_{\scriptscriptstyle B} \ll T^2$. This shows that, the smaller the temperature $T$, the weaker the coupling needs to be in order for the perturbative calculations to make sense. An analogue of Eq.~\eqref{eq:elements2} was derived in Ref.~\cite{thingna2012} and for the special case of linear coupling to a bosonic bath in Ref.~\cite{cresser2021weakandultrastrong}.

Thus equipped with Eqs.~\eqref{eq:tay1} and \eqref{eq:elements2}, we can turn to Eq.~\eqref{eq:lyapunov} and write the SLD as
\begin{subequations}\label{eq:SLD}
\begin{align}
\logd(\beta)=-\Delta\hams+\gamma^2\sum\nolimits_{n m} \alpha_{nm}(\beta)\ket{\epsilon_n}\bra{\epsilon_m}+O(\gamma^4)
\end{align}
with $\Delta \hams \coloneqq \hams - \avs{\hams}$ and coefficients
\begin{align}\label{alpha_nm}
\alpha_{nm}(\beta)=\frac{2\, p_n}{p_n+p_m}\, \partial_\beta (\Xs)_{nm} + \frac{p_n \Delta_{nm}}{p_n+p_m}\, (\Xs)_{nm}.
\end{align}    
\end{subequations}
Here, $ \Delta_n \coloneqq \epsilon_n-\avs{\hams} $. Also note that the populations appearing above are 
\begin{equation}
    p_n = \bra{\epsilon_n} \pis \ket{\epsilon_n} = e^{-\beta\epsilon_n}/\parts
\end{equation}
and should not be confused with the populations of the mean-force Gibbs state 
\begin{equation}\label{eq:poppist}
    \tilde{p}_n = \bra{\epsilon_n}\pist\ket{\epsilon_n} = p_n + \gamma^2 p_n (\Xs)_{nn} +O(\gamma^4).
\end{equation}
We now have the two ingredients needed to compute the QFI from Eq.~\eqref{eq:defqfi}. After some algebra, this can be compactly written as
\begin{align}\label{eq:result2}
\fish=\partial^2_\beta\ln \parts+\gamma^2\int^{\beta}_0 du \ \left[\tilde{\Phi}_{\scriptscriptstyle B}(-iu)\, \partial^2_\beta\Phi_{\scriptscriptstyle S}(-iu)
+2\, \partial_\beta\tilde{\Phi}_{\scriptscriptstyle B}(-iu)\, \partial_\beta\Phi_{\scriptscriptstyle S}(-iu)\right]+O(\gamma^4), 
\end{align}
with the modified auto-correlation function $\tilde{\Phi}_{\scriptscriptstyle B}(x)\coloneqq(\beta-ix)\Phi_{\scriptscriptstyle B}(x)$. To second order in $\gamma$ the QFI can thus be expressed solely in terms of the probe and sample correlation functions. We emphasise that the validity of Eqs.~\eqref{eq:SLD} and~\eqref{eq:result2} is not, in any way, underpinned by the Born--Markov approximation.

Equations~\eqref{eq:SLD} and~\eqref{eq:result2} are our second main result. They can be applied directly to \textit{any} open quantum system under finite coupling, provided that Assumptions I and II are met. And since the systems satisfying these assumptions span an exceptionally wide range, our closed-form expressions for $\logd$ and $\fish$ open up new avenues in quantum thermometry. Note that, when Assumptions I and II are violated, the QFI will in general pick up a first-order term, as can be anticipated from Eq.~\eqref{eq:firstorder}. We provide illustrative examples of that in Appendix~\ref{app:counterexamples}.

Finally, it is often useful to talk in terms of $ T $ rather than inverse temperature $ \beta $. Namely, from Eq.~\eqref{eq:cramer-rao} we see that the best-case signal-to-noise ratio (SNR) for estimates of $ T $ can be cast as
\begin{align}
\frac{1}{N}\left(\frac{T}{\delta T}\right)^2\leq\beta^2\, \fish \coloneqq C_{\scriptscriptstyle S}(T)+\gamma^2 \xi(T), \label{correction}
\end{align}
with $C_{\scriptscriptstyle S}(T) = \beta^2\partial^2_\beta\ln\parts = \partial_T\avs{\hams} $ being the heat capacity of the probe when in local thermal equilibrium \cite{Correa2015}, and where $\xi(T)$ follows from the second-order term in Eq.~\eqref{eq:result2}.

\section{Examples}
\label{sec:examples}

We now benchmark our Eq.~\eqref{eq:result2} against the exactly solvable Caldeira--Leggett model for quantum Brownian motion and then, apply it to the spin--boson problem. In both cases, we model the sample as an infinite collection of uncoupled harmonic oscillators spanning a quasi-continuum of frequencies---the most common bath model in open quantum systems. That is, 
\begin{equation}\label{eq:hamb_specific}
    \hamb = \sum\nolimits_k \omega_k\, \op{b}^\dagger_k \op{b}_k, 
\end{equation}
where $\op{b}_k^\dagger$ ($\op{b}_k$) is the creation (annihilation) operator at frequency $ \omega_k $. As it is also customary, we take the coupling operator $ \op{B} $ to be
\begin{equation}\label{eq:opB_specific}
    \op{B} = \sum\nolimits_k \left(g_k\, \op{b}_k^\dagger + g_k^*\, \op{b}_k\right), 
\end{equation}

Here, the coupling constants $ g_k $ encode the overall dissipation strength (note that $\gamma$ still enters $\op{H}_\text{int}$). We may collect them into the \textit{spectral density} function
\begin{equation}
    J(\omega) = \pi\, \sum\nolimits_k \vert g_k \vert^2\, \delta(\omega-\omega_k), 
\end{equation}
where $\delta(\cdot)$ stands here for the Dirac delta. Since we assume that the sample has a quasi-continuum spectrum, we may give some smooth functional form to $ J(\omega) $. 

Using this notation, the bath correlation function evaluates to
\begin{align}
\Phi_{\scriptscriptstyle B}(-iu) = \frac{1}{\pi}\int^\infty_0 d\omega \ J(\omega) \, \frac{\cosh{\left(\beta\, \omega/2-u\, \omega\right)}}{\sinh{\left(\beta\, \omega/2\right)}}.
\end{align}
Furthermore, using Eq.~\eqref{eq:result2} we see that the second-order correction $\xi(T)$ to the SNR from Eq.~\eqref{correction} is given by
\begin{subequations}\label{eq:correction-normal-bath}
\begin{align}
\xi(T)= \int^\infty_0 d\omega \ J(\omega)\, f_{\scriptscriptstyle S}(T, \omega) \label{eq:int}, 
\end{align}
where 
\begin{align} \nonumber
f_{\scriptscriptstyle S}(T, \omega)=&\frac{\beta^2}{\pi \sinh{\frac{\beta\, \omega}{2}}}\int^\beta_0 du \ (\beta-u)\, \cosh{\left(\beta\, \omega/2-u\, \omega\right)}
\\ \label{eq:fs}
&\times \Big[\partial^2_\beta\Phi_{\scriptscriptstyle S}(-iu)+\partial_\beta\Phi_{\scriptscriptstyle S}(-iu)\Big(\frac{2}{\beta-u}-\omega\coth\frac{\beta\, \omega}{2} - \omega\tanh\frac{(2u - \beta) \, \omega}{2}\Big) \Big] .
\end{align}    
\end{subequations}

Eqs.~\eqref{eq:correction-normal-bath} above hold for any spectral density $ J(\omega) $, provided that the bath and the probe--bath coupling are of the standard type defined by Eqs.~\eqref{eq:hamb_specific} and \eqref{eq:opB_specific}. Hence, Eq.~\eqref{eq:correction-normal-bath} holds even where most master equations break down, e.g., when the bath correlation function happens to be long lived. As we illustrate below, Eqs.~\eqref{eq:correction-normal-bath} are indeed very practical to quickly evaluate the finite-coupling corrections to thermal sensitivity, and provide analytic intuition about their temperature dependence. This is our third result.

Specifically, in what follows we work with a spectral density with variable Ohmicity $ s $ and exponential cutoff; namely, 
\begin{align}\label{eq:spec}
    J(\omega) = \Omega^{a-s}\, \omega^s\, e^{-\omega/\Omega},
\end{align}
where $a$ controls the dimension of $J(\omega)$ and therefore depends on the choice of $\op{S}$: if [$\op{S}$]$=$[$\omega$]$^\mu$, then $a = 1 - 2 \mu$, where [$\cdot$] gives the dimension of the argument.

\subsection{Quantum Brownian motion} 

In order to put the accuracy of our second-order formula to the test, we take a single harmonic oscillator as the probe. That is, 
\begin{align}
\hams= \frac{1}{2} \omega_0^2 \op{x}^2 + \frac{1}{2} \op{p}^2, 
\end{align}
while the probe operator entering in the interaction Hamiltonian\footnote{Note that we are not adding a counter-term to compensate for the distortion of the harmonic potential due to the coupling to the sample. While it may be important to do so whenever such renormalisation is not expected to be physical---especially when the coupling is strong---this effect is irrelevant for us.} in Eq.~\eqref{eq:Hint} is $\op{S} = \op{x}$. For this choice of $\op{S}$, dimensional analysis imposes the value $a = 2$ for the exponent in Eq.~\eqref{eq:spec}. {\blue The thermal sensitivity may be computed using the explicit formula in Eqs. (11) and (12) in \cite{Mehboudi2019}. In turn, exact expressions for the steady-state covariances of this system are given in Appendices B--E of \cite{correa2017thermometry_strongcoupling}. This allows us to benchmark Eq.~\eqref{eq:result2}.} 

The only ingredient missing from Eqs.~\eqref{eq:correction-normal-bath} is the probe correlation function, for which a straighforward calculation yelds
\begin{align}
    \Phi_{\scriptscriptstyle S}(-iu)=\frac{1}{2 \omega_0}  e^{-u\, \omega_0} + \frac{1}{\omega_0} \frac{\cosh{(u\, \omega_0)}}{e^{\beta\, \omega_0}-1}.
\end{align}

\begin{figure}[t]
\centering
\includegraphics[width=0.6\textwidth]{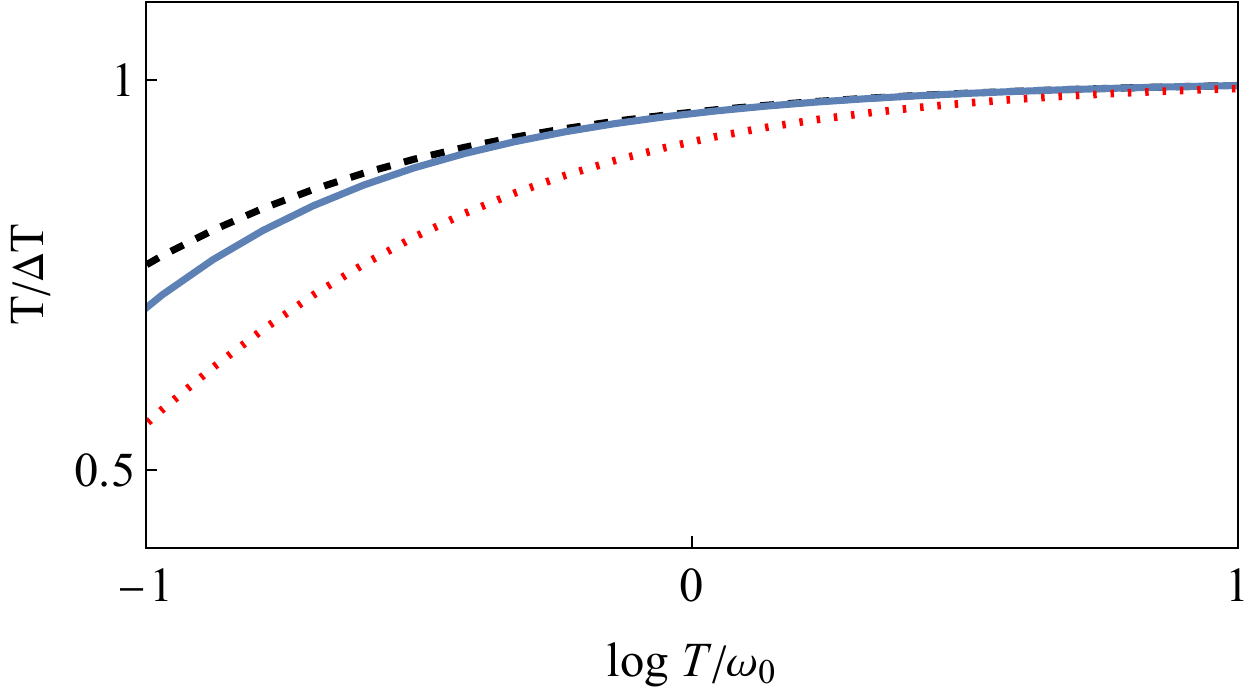}
\caption{\textbf{Benchmarking Eq.~\eqref{eq:result2} for quantum Brownian motion.} The plot shows the signal-to-noise ratio from our second-order formula (solid blue), the true SNR ratio from the exact mean-force Gibbs state (dashed black), and from the infinitesimal coupling limit, i.e. the local Gibbs state (dotted red). The second-order formula tracks the exact sensitivity well as long as the probe--sample coupling is the only small parameter. It starts to break down when the temperature itself becomes a small parameter as $T \rightarrow 0$. We chose $\omega_0=1, \gamma=0.1, s=1$ and $\Omega=100$ ($\hbar=k=1$).}
\label{fig1}
\centering
\end{figure}

In Fig.~\ref{fig1} we compare the signal-to-noise ratio according to Eq.~\eqref{correction} against the exact result and the thermal sensitivity of a probe in local thermal equilibrium, which is given by the heat capacity alone. We see that our second-order formula accurately captures the sensitivity for small coupling. As already advanced, at very low temperatures, $\beta$ itself sets an energy-scale compared to which the coupling is no longer perturbative. This results in our second-order approximation becoming less accurate. On the other hand, the sensitivity saturates at high temperature, and is correctly approximated by the assumption of local thermal equilibrium on the probe, as expected. Hence, our formula provides substantial quantitative advantage in the range of moderately low temperatures.

To get further insight into the correction we calculate the high-temperature expansion of Eq.~\eqref{eq:fs}, which gives
\begin{align}
f_{\scriptscriptstyle S}(T, \omega) = - \frac{\beta^2}{6 m \pi \omega} + O(\beta^4), 
\end{align}
doing the integral in Equation~\eqref{eq:int} for our spectral density \eqref{eq:spec} gives
\begin{align}
\xi(T)=-\frac{\Omega^2 \, \Gamma\left( s \right) }{3 m \pi}\, \beta^2 + O(\beta^4), 
\end{align}
with $\Gamma\left( \cdot \right)$ the Euler gamma function. This is exactly the lowest-order contribution from the frequency renormalisation due to the finite interaction with the sample, i.e. the classical mean-force correction \cite{cerisola2022qcc}.

\subsection{Spin--boson model}\label{sec:spin-boson} 

\begin{figure}[t]
\centering
\includegraphics[width=0.53\textwidth]{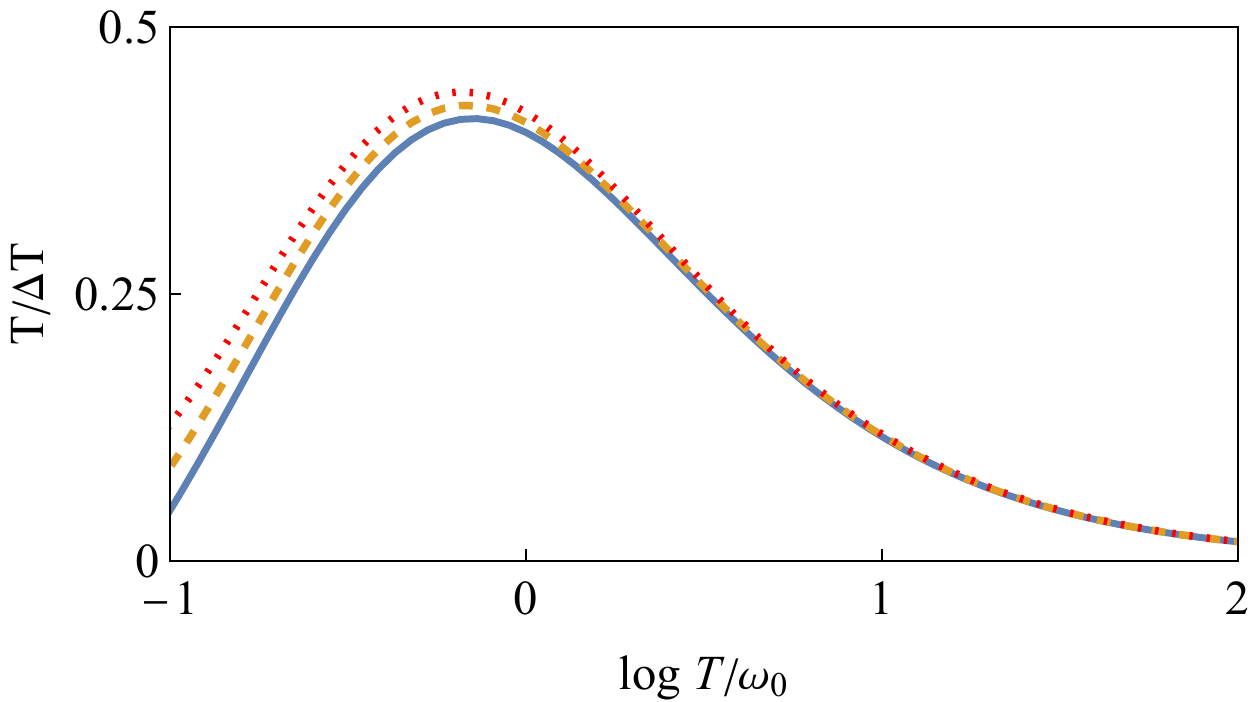}
\includegraphics[width=0.46\textwidth]{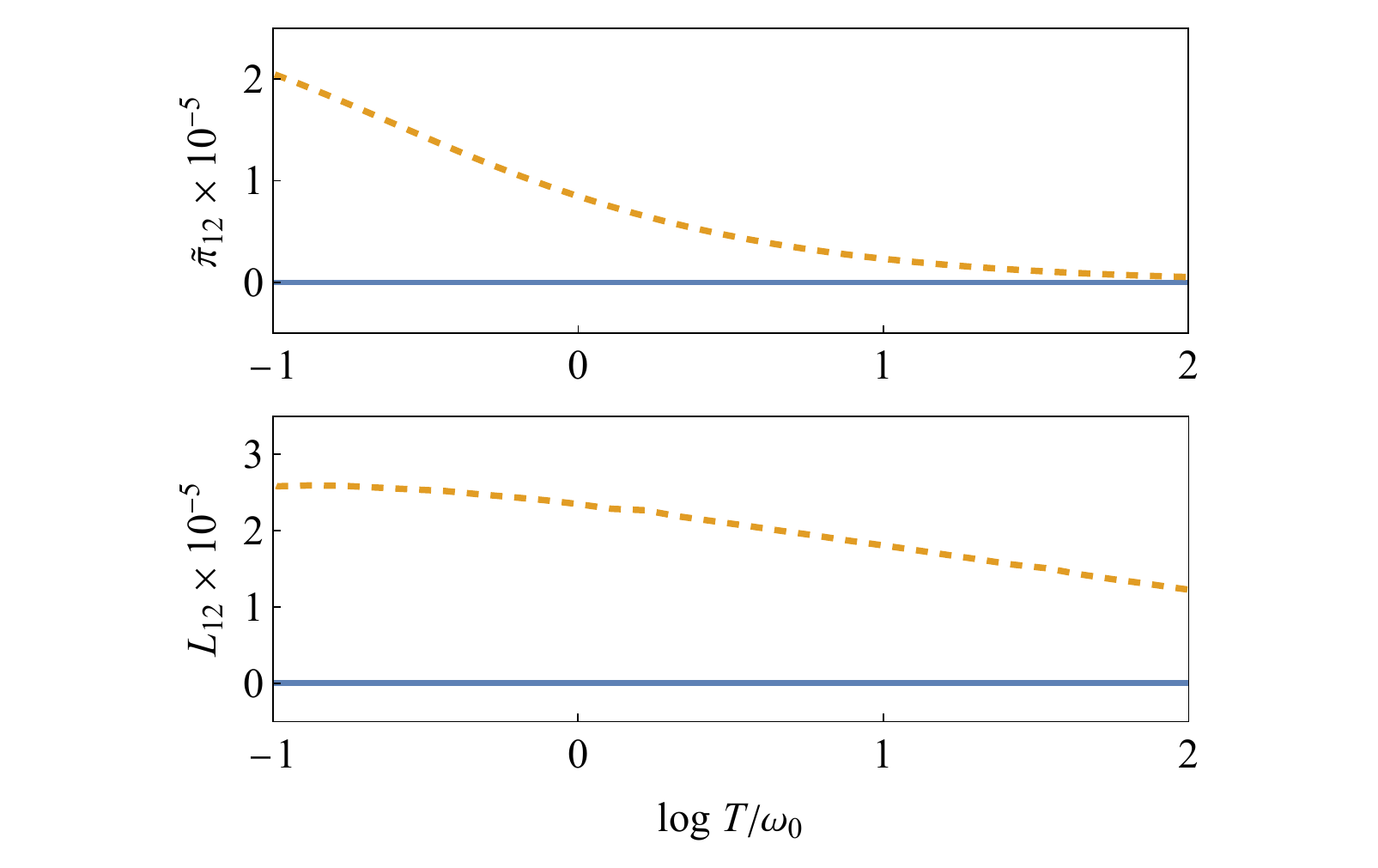}
\caption{\textbf{Applying Eq.~\eqref{eq:result2} to the spin--boson model.} The left panel shows the SNR from our second-order formula for two types of probe--sample coupling; namely, $\theta=3\pi/2$, i.e. $\op{S}=\op{\sigma}_x$ (solid blue) and for $\theta=\pi/4$, i.e. the maximally coherent case (dashed orange). We compare these to the sensitivity of a locally thermal probe (dotted red). The top-right panel shows the second-order coherences of the respective mean-force Gibbs states. The same color coding applies. The bottom-right plot shows the second-order coherences of the respective symmetric logarithmic derivatives in the basis of $\hams$. We emphasise that diagonal energy measurements saturate the second-order precision bound despite $\pist$ and $\logd$ acquiring temperature-dependent off-diagonal elements at second order in $\gamma$. The other parameters are the same as in Fig.~\ref{fig1}.}
\label{fig2}
\centering
\end{figure}

We now apply our formula to a case study for which no exact solution is available---the spin--boson model. In order to study the effects of coherence in the steady state, we focus on the `$\theta$-angled' spin--boson model. That is, 
\begin{subequations}
    \begin{align}
\hams&=\frac{\epsilon}{2}\op{\sigma}_z, \\
\op{S}&=\cos{\theta}\, \op{\sigma}_z - \sin{\theta}\, \op{\sigma}_x, 
\end{align}
\end{subequations}
where $\op{\sigma}_\alpha$ are Pauli operators. For this choice of $\op{S}$, $a = 1$. We thus have the probe correlation function
\begin{align}
    \Phi_{\scriptscriptstyle S}(-iu)=\cos^2{\theta}+\cosh{\left(\epsilon\, u - \beta\, \epsilon/2\right)}\, \sech{\left(\beta\, \epsilon/2\right)}\, \sin^2{\theta}.
\end{align}
In Fig.~\ref{fig2} we compare the the SNR ratio as per Eq.~\eqref{correction} for two values of $\theta$ and, once again, also against the sensitivity of a locally thermal probe. We also plot the coherence appearing in our approximation of the mean-force Gibbs state [Eq.~\eqref{eq:tay1}], and the coherence showing up in the symmetric logarithmic derivative, to second order in $\gamma$ [Eq.~\eqref{eq:SLD}]. In this case, however, there is no exact SNR to compare with. We see that the two different couplings yield slightly different sensitivity, departing from the SNR of the locally thermal probe. We can also see that, at large enough temperature, the sensitivity drops to zero, as the spin populations saturate.

Rather counter-intuitively, we can see in the right panel that the $\theta=\pi/4$ coupling generates \textit{temperature-dependent} coherences in the second-order approximation to $\pist$, and in the second-order corrections to $\logd$. Yet, we have proven that (diagonal) measurements in the local energy basis $\hams$---which destroy such temperature information---do still saturate the ultimate precision bound, at least \textit{to second order} in $\gamma$ [cf. Eq.~\eqref{eq:resultmain}]. In other words, temperature-dependent coherences may only improve precision at higher orders in finite-coupling quantum thermometry. 

It is interesting to note that, in this case, the correction $\xi(T)$ to the optimal SNR has a definite sign---it is always negative. Indeed, the factor in brackets in Eq.~\eqref{eq:fs} is negative for all $\omega$, while the spectral density is obviously positive for all $\omega$. Hence, the weaker the probe--sample coupling the better for this type of probe. For completeness, we calculate the high--$T$ limit of Eq.~\eqref{eq:fs}. We get
\begin{align}
f_{\scriptscriptstyle S}(T, \omega)=-\frac{2 \sin^2\left( \theta \right) \epsilon^2 }{3 \pi \omega}\, \beta^3 + O(\beta^5)
\end{align}
which, after integration as per Eq.~\eqref{eq:int}, yields
\begin{align}
\xi(T)=-\frac{4 \sin^2\left( \theta \right) \epsilon^2 
\Omega \Gamma\left( s \right) }{3 \pi}\, \beta^3 + O(\beta^5).
\end{align}
This highlights what can already be seen in Fig.~2---that here, the vanishing-coupling limit gives a far better approximation to the thermal sensitivity at moderate-to-large temperature, when compared with the quantum Brownian motion example.

\section{Derivation}\label{sec:derivation}

\subsection{Mean-force Gibbs state}

In this section we elaborate on the derivation of the results presented in Sec.~\ref{sec:qt}. The first step in our search for an explicit expression for $ \fish $ will be to obtain a second-order approximation of the reduced probe state $\pist$, following \cite{thingna2012}. Applying Eq.~\eqref{lem:1} to $e^{-\beta \op{H}}$ leads to:
\begin{align}\label{eq:exp1}
\nonumber e^{-\beta \op{H}}=&e^{-\beta(\hams+\hamb)}\bigg(\id-\gamma \int^\beta_0 d\beta_1 \ e^{\beta_1 (\hams+\hamb)} \op{V} e^{-\beta_1 (\hams+\hamb)} \\
&+\gamma^2 \int^\beta_0 d\beta_1 \ \int^{\beta_1}_0 d\beta_2 \ e^{\beta_1 (\hams+\hamb)} \op{V} e^{-(\beta_1-\beta_2) (\hams+\hamb)} \op{V} e^{-\beta_2 (\hams+\hamb)} \bigg)+O(\gamma^3).
\end{align}
Taking the trace yields an approximation for the total partition function
\begin{multline}
\tilde{Z} = \parts\partb-\gamma\, \beta\, \trS{e^{-\beta\hams}\op{S}}\trB{e^{-\beta\hamb}\op{B}} \\
+\gamma^2\int^\beta_0 d\beta_1 \ \int^{\beta_1}_0 d\beta_2 \ \trS{e^{-\beta\hams}e^{(\beta_1-\beta_2)\hams}\op{S}\, e^{-(\beta_1-\beta_2)\hams}\op{S}}\\ \times\trB{e^{-\beta\hamb}e^{(\beta_1-\beta_2)\hamb} \op{B}\, e^{-(\beta_1-\beta_2)\hamb} \op{B}}+O(\gamma^3),
\end{multline}
where Assumption I [Eq.~\eqref{eq:Hint}] has been applied. Using $(1+x)^{-1} = 1-x+O(x^2)$, Assumption II [Eq.~\eqref{eq:avgB}], and the definition of probe and sample auto-correlation functions from Eq.~\eqref{eq:correl}, we find 
\begin{align}\label{eq:part}
\frac{1}{\tilde{Z}}=\frac{1}{\parts\partb}\bigg(1-\gamma^2\int^\beta_0 d\beta_1 \ \int^{\beta_1}_0 d\beta_2 \ \Phi_{\scriptscriptstyle S}(-i(\beta_1-\beta_2))\Phi_{\scriptscriptstyle B}(-i(\beta_1-\beta_2))\bigg)+O(\gamma^4), 
\end{align}
Multiplying Eqs.~\eqref{eq:exp1} and~\eqref{eq:part}, collecting terms, and taking the partial trace over the sample, we find that the first order correction vanishes, so that
\begin{align*}
\pist=\frac{1}{\tilde{Z}}\trB{e^{-\beta \op{H}}}=\pis\big(\ids+\gamma^2\Xs\big)+O(\gamma^4), 
\end{align*}
where we have introduced the operator
\begin{multline}\label{eq:X}
\Xs=\int^{\beta}_0 d\beta_1 \ \int^{\beta_1}_0 d\beta_2 \ \Phi_{\scriptscriptstyle B} (-i(\beta_1-\beta_2)) \\ \times \big[e^{\beta_1\hams} \op{S} e^{-(\beta_1-\beta_2))\hams} \op{S} e^{-\beta_2\hams}-\Phi_{\scriptscriptstyle S}(-i(\beta_1-\beta_2))\ids \big], 
\end{multline}
Note that $\trS{\Xs\pis}=0$ as required by normalisation.

Writing the diagonal elements of the operator $(\Xs)_{nn} \coloneqq \bra{\epsilon_n}\Xs\ket{\epsilon_n}$ in the eigenbasis of $\hams=\sum\nolimits_n \epsilon_n \ket{\epsilon_n}\bra{\epsilon_n}$ gives
\begin{multline}\label{eq:elements}
(\Xs)_{nn}=\int^{\beta}_0 d\beta_1 \ \int^{\beta_1}_0 d\beta_2 \ \Phi_{\scriptscriptstyle B}(-i(\beta_1-\beta_2))\\ \times\big[\sum\nolimits_k \phi_{nk}(-i(\beta_1-\beta_2))-\Phi_{\scriptscriptstyle S}(-i(\beta_1-\beta_2))\big], 
\end{multline}
which can be turned into Eq.~\eqref{eq:elements2} by introducing the new variables $u=\beta_1-\beta_2$ and $v=\beta_1+\beta_2$ and performing the integral over $v$; namely
\begin{align*}
(\Xs)_{nn}&= \frac12\, \int_0^\beta du\, \int_u^{2\beta-u} dv\, \Phi_{\scriptscriptstyle B}(-iu)\big(\sum\nolimits_k \phi_{nk}(-iu)-\Phi_{\scriptscriptstyle S}(-iu)\big)\nonumber\\ &=\int^{\beta}_0 du \ (\beta-u)\Phi_{\scriptscriptstyle B}(-iu)\big(\sum\nolimits_k \phi_{nk}(-iu)-\Phi_{\scriptscriptstyle S}(-iu)\big), 
\end{align*}
In turn, the off-diagonal elements of $ \Xs $ are
\begin{align}
(\Xs)_{nm}=\sum\nolimits_k (\op{S})_{nk}(\op{S})_{km}\int^\beta_0 d\beta_1 \ e^{\beta_1 \Delta_{kn}}\int^{\beta_1}_0 d\beta_2 \ e^{\beta_2 \Delta_{mk}} \Phi_{\scriptscriptstyle B}(-i(\beta_1-\beta_2))
\end{align}
Again, we may introduce $u=\beta_1-\beta_2$ and $v=\beta_1+\beta_2$ and integrate over $v$, which gives
\begin{align}\label{eq:elements4}
(\Xs)_{nm}=\frac{1}{\Delta_{mn}}\sum\nolimits_k (\op{S})_{nk}(\op{S})_{km}\, \int_0^\beta du\, \Phi_{\scriptscriptstyle B}(-iu)\, \left( e^{u\Delta_{km}}\, e^{\beta \Delta_{mn}}-e^{u\Delta_{kn}} \right) \ \ \ \ (n\neq m).
\end{align}

\subsection{Symmetric logarithmic derivative}

We now use the expression for the mean-force Gibbs state to obtain the SLD via Eq.~\eqref{eq:lyapunov}. To do so, we differentiate Eq.~\eqref{eq:tay1} and use $\avs{\hams}=-\partial_\beta \ln \parts$ to get
\begin{align}\label{eq:tay2}
\partial_\beta\, \pist=-\Delta \hams\, \pis+\gamma^2\partial_\beta(\pis\Xs)+O(\gamma^4).
\end{align}
Turning to the operator $e^{-\lambda \pist}$ and combining Eqs.~\eqref{lem:1} and \eqref{eq:tay1} \blue{and setting $\xi = \gamma^2$} leads to\footnote{\blue{Note that lifting Assumption II would leave us with $\tilde{\pmb{\pi}}_S = \pmb{\pi}_S + \gamma\pmb{p}_1 + \gamma^2\pmb{p}_2$. Hence, Eqs.~\eqref{lem:1} would need to be suitably generalised to handle the expansion of $ e^{-\lambda(\pmb{\pi}_S+\gamma\pmb{p}_1+\gamma^2\pmb{p}_2)} $, for substitution into \eqref{eq:lyapunov}.}} 
\begin{align}\label{eq:tay3}
e^{-\lambda \pist}=e^{-\lambda \pis}\left(\ids-\gamma^2\pis\int^\lambda_0 d\lambda' \ e^{\lambda'\pis}\Xs \, e^{-\lambda'\pis}\right)+O(\gamma^4).
\end{align}
Using this, we can finally compute the SLD. From Eqs.~\eqref{eq:tay2} and~\eqref{eq:tay3} we get
\begin{multline}\label{eq:SLDtay}
e^{-\lambda \pist}(\partial_\beta\, \pist)e^{-\lambda \pist}=-\Delta\hams\, \pis e^{-2\lambda \pis}
\\
+\gamma^2 \pis \int^\lambda_0 d\lambda' \ e^{-(\lambda-\lambda') \pis}\Xs\, e^{-(\lambda+\lambda') \pis} \Delta\hams\, \pis,
\\
+\gamma^2 e^{-\lambda \pis}\partial_\beta (\pis\Xs)\, e^{-\lambda \pis}
\\
+\gamma^2\Delta\hams\, \pis^2 e^{-2\lambda \pis}\int^\lambda_0 d\lambda' \ e^{\lambda'\pis}\Xs\, e^{-\lambda' \pis}+O(\gamma^4)
\end{multline}
We can now recover Eq.~\eqref{eq:SLD} using Eqs.~\eqref{eq:lyapunov} and working again in the eigenbasis of $\hams$. Namely, 
\begin{align*}
\logd(\beta)=-\Delta\hams+\gamma^2\sum\nolimits_{n m} \alpha_{nm}(\beta)\ket{\epsilon_n}\bra{\epsilon_m}+O(\gamma^4), 
\end{align*}
with coefficients
\begin{align*}
\alpha_{nm}(\beta)=\frac{2\, p_n}{p_n+p_m}\, \partial_\beta (\Xs)_{nm} + \frac{p_n \Delta_{nm}}{p_n+p_m}\, (\Xs)_{nm}.
\end{align*} 
Here, we have used the fact that $\partial_\beta\, p_n = p_n\, (\langle H_{\scriptscriptstyle S}\rangle_{\scriptscriptstyle S}-\epsilon_n) = -p_n\, \Delta_n$ (not to be confused with the energy gaps $\Delta_{nk}$).
As long as we know the sample correlation function $\Phi_{\scriptscriptstyle B}(x)$ and the eigenstates of $\hams$, we will be able to obtain the elements $(\Xs)_{nm}$ from Eqs.~\eqref{eq:elements2} and \eqref{eq:elements4}, and thus, the SLD as per Eq.~\eqref{eq:SLD}. 

\subsection{Quantum Fisher information}

The QFI can now be directly computed by using $\fish=\trS{\logd^2\, \pist}$. Collecting terms up to second order yields
\begin{align}\label{eq:result}
\fish = \partial^2_\beta\ln \parts + \gamma^2\sum\nolimits_{n}\left[p_n \Delta_n^2 (\Xs)_{nn} - 2 \, p_n \Delta_n \alpha_{nn} \right] + O(\gamma^4), 
\end{align} 
where $\partial^2_\beta\ln \parts = \langle(\Delta\hams)^2\rangle_S$ and $\alpha_{nn} = \partial_\beta(\Xs)_{nn}$, by Eq.~\eqref{alpha_nm}. 

We can still obtain the much more elegant expression \eqref{eq:result2}, in terms of probe and sample auto-correlation functions alone. Noting that
\begin{align*}
    \Phi_S(-iu) = \sum\nolimits_{nk} \phi_{nk} (-iu)
\end{align*}
is easy to see that
\begin{subequations}
\begin{align}
    \sum\nolimits_n p_n\Delta_n(\Xs)_{nn} &= -\partial_\beta\Phi_{\scriptscriptstyle S}(-iu) \\
    \sum\nolimits_n p_n\Delta_n^2 (\Xs)_{nn} &= \partial_\beta^2\Phi_{\scriptscriptstyle S}(-iu). 
\end{align}
\end{subequations}

Furthermore, since $\sum\nolimits_n p_n\, \Delta_n = 0$, we see that $\partial_\beta(\Xs)_{nn}$ reduces to
\begin{align}
    \partial_\beta(\Xs)_{nn} = - \partial_\beta\tilde{\Phi}_{\scriptscriptstyle B}(-iu)\, \partial_\beta\Phi_{\scriptscriptstyle S}(-iu), 
\end{align}
where, recall that $ \tilde{\Phi}_{\scriptscriptstyle B}(-iu) = (\beta-u)\, \Phi_{\scriptscriptstyle B}(-iu) $. Hence, Eq.~\eqref{eq:result} can be recast as desired:
\begin{align*}
\fish=\partial^2_\beta\ln \parts+\gamma^2\int^{\beta}_0 du \ \left[\tilde{\Phi}_{\scriptscriptstyle B}(-iu)\, \partial^2_\beta\Phi_{\scriptscriptstyle S}(-iu)
+2\, \partial_\beta\tilde{\Phi}_{\scriptscriptstyle B}(-iu)\, \partial_\beta\Phi_{\scriptscriptstyle S}(-iu)\right]+O(\gamma^4).
\end{align*}

\subsection{Classical Fisher information for energy measurements}

Finally, we connect with our general result in Eq.~\eqref{eq:resultmain}. To that end, we use the definition \eqref{eq:defcfi} of the (classical) Fisher information associated to energy measurements of $\pist$ on the basis of the bare Hamiltonian $\hams$. Namely, 
\begin{align}
\cfish = \sum\nolimits_n \tilde{p}_n\, (\partial_\beta \ln \tilde p_n)^2.
\end{align}
Plugging in the mean-force Gibbs state populations from Eq.~\eqref{eq:poppist} gives
\begin{align*} \nonumber
\cfish &= \sum\nolimits_n \tilde p_n (\partial_\beta \ln \tilde p_n)^2
\\
&= \partial_\beta^2\ln{\parts}+\gamma^2\sum\nolimits_n \left[p_n \Delta_n^2 (\Xs)_{nn}-2 \Delta_n  \partial_\beta (\Xs)_{nn}\right] + O(\gamma^4).
\end{align*}
Here, we have used the identities $\ln \tilde p_n = p_n + \gamma^2 (\Xs)_{nn} +O(\gamma^4)$, $\partial_\beta \ln p_n = - \Delta_n$, and the fact that $\sum\nolimits_n p_n (\partial_\beta \ln p_n)^2= \partial_\beta^2 \ln \parts$. As expected, we thus recover Eq.~\eqref{eq:result}, meaning that local energy measurements on the probe are as efficient as the truly optimal measurements in finite-coupling quantum thermometry. This statement holds generally to second order in the coupling for any thermometric setup. The only underpinning assumption is that probe and sample have a separable coupling, and that the sample operator averages to zero when in a local thermal state. This choice for the dissipative interaction is indeed widespread in open quantum systems.

\section{Conclusions}\label{sec:conclusions}

We have developed a theory of finite-coupling quantum thermometry, focusing on the second-order correction terms. On the one hand, we have obtained closed-form expressions for the symmetric logarithmic derivative, which fixes the optimal basis in which to measure the probe; and the quantum Fisher information, which gauges the responsiveness of the probe to temperature fluctuations on the sample. Our formulas make minimal assumptions on the probe--bath coupling and can be readily applied to any setup by simply computing the probe and sample auto-correlation functions. In particular, we make no assumptions on the nature of the sample, so that our framework is applicable even when standard open-system weak-coupling tools---based on the Born--Markov approximation---break down.

On the other hand, we have proved that \blue{the sensitivity of} projective measurements in the local energy basis of the probe is always optimal to the first two leading orders in the probe--sample coupling. That is, even if temperature information may be encoded in the off-diagonal elements of the probe's marginal state, the advantage in extracting and processing such information would only show up in higher order terms. Hence, measurement optimisation is unnecessary for quantum thermometry at moderate couplings.

Finally, we have illustrated the use of our approach with two relevant examples---the exactly solvable Caldeira--Legett model (which served as benchmark) and the spin--boson model. Specifically, we have derived an explicit expression for the second-order correction to the signal-to-noise ratio for temperature estimation, suitable for any open-system model which uses a canonical linear bath with quasi-continuum spectrum; regardless of the specific spectral density. Such formula can thus be of independent interest.

Our results open up new directions in finite-coupling thermometry. Namely, our formulas can easily deal with thermometry on more realistic non-linear sample models, which remain virtually unexplored to date. They can also facilitate the treatment of complex probe--sample interactions, such as the ones appearing when modelling an atomic impurity in a Bose--Einstein condensate \cite{Mehboudi2019}, without the need for \textit{linearisation}. Crucially, having established the optimality of \blue{the sensitivity of} energy measurements up to second order in the coupling strength highlights the importance of directing practical efforts towards measuring the probe in its local basis. Furthermore, guaranteeing that optimal \blue{thermometric sensitivity} may be attained using a fixed \textit{temperature-independent} basis also eliminates the need for more complex adaptive schemes.

Interestingly, lifting our Assumption II on the vanishing expectation value of the interaction in the stationary state of the sample would result in inhomogeneous terms, which might introduce a thermometric advantage in terms of additional temperature dependence. Studying how this may be exploited in practice will be the subject of future study.

\blue{Lastly, we emphasise that equilibrium thermometry assumes that the probe and sample are (jointly) thermalised. In practice, however, this assumption can break down whenever the thermalisation time is comparable to the reset time of the experiment. In those cases, the transient probe--sample dynamics becomes important. Such non-equilibrium thermometry has been studied \cite{johnson2016thermometry,mitchison2020situ}, often in the framework of weak-coupling master equations \cite{guo2015ring, Feyles2017dynamical, Sekatski2022}. An alternative approach to dynamical thermometry is the use of collisional ancilla-systems as probe \cite{seah2019collisional, Kiilerich2018dynamical, pati2020weak, boeyens2021noninformative, boeyens2023probe}. In presence of periodic driving even the long time behaviour becomes time dependent, as the probe settles into a limit cycle instead of a steady state \cite{mukherjee2019thermometry_control, glatthard2022bending}. Examining the dynamical implications of the finite probe--sample coupling scenario studied here remains an open problem that could be tackled using methods developed in the context of quantum control \cite{kofman2000acceleration, Kofman2004unified, erez2008control, kurizki2022thermo}.}

\acknowledgements

JG is supported by a scholarship from CEMPS at the University of Exeter. LAC acknowledges the Spanish Ram\'{o}n y Cajal program (Fellowship: RYC2021-325804-I) funded by MCIN/AEI/10.13039/501100011033 and “NextGenerationEU”/PRTR. KVH acknowledges support by the University of Potsdam startup funds. MPL acknowledges support by the the Swiss National Science Foundation through an Ambizione Grant No. PZ00P2-186067. HM acknowledges support from the Royal Commission for the Exhibition of 1851. LAC, KVH, MPL and HM acknowledge the US National Science Foundation (Grant No. NSF PHY1748958) and thank the Kavli Institute for Theoretical Physics for their warm hospitality during the program \textit{``Thermodynamics of quantum systems: Measurement, engines, and control''}, during which this (long overdue) project started.

\appendix

\section{Examples with $\avb{\op{B}}\neq 0$}
\label{app:counterexamples}

In this short Appendix, we give explicit examples in which Assumption II in Eq.~\eqref{eq:avgB}---namely $\avb{\op{B}} = 0 $---is not satisfied. Specifically, we \blue{illustrate} how, in this case, the QFI and SLD may pick up terms of $O(\gamma)$.

\subsection{\blue{Non-dissipative coupling}}
\label{app:counterexamples_A}

First, let us consider the two-qubit model
\begin{equation*}
\op{H} = \hams + \hamb + \op{H}_\text{int} = \omega\,\op{\sigma}_{z,1}\otimes\id_2 + \omega\,\id_1\otimes\op{\sigma}_{z,2} + \gamma\,\op{\sigma}_{z,1} \otimes \op{\sigma}_{z,2},   \end{equation*}
where the sub-indices label the qubits, `1' being the probe ($S$) and `2' being the sample ($B$). Also, in all the expressions that follow, we shall set $ \omega = 1 $ for simplicity.

The corresponding mean-force Gibbs state can be expanded as
\begin{equation*}
    \pist = \pis + \gamma\,p_1\,\op{\sigma}_{z,1} + \gamma^2 p_2\,\op{\sigma}_{z,1} + O(\gamma^3),
\end{equation*}
with $p_1 = \frac12\,\beta\,\sech^2{\beta}\,\tanh{\beta}$ and $p_2 = \frac12\,\beta^2\,\sech^2{\beta}\,\tanh^3{\beta}$. As we anticipated, in this case the QFI does feature a linear term, namely
\begin{equation*}
    \fish = \partial_\beta^2\ln{Z_S} + \sech^4{\beta} \left[\beta  (\cosh{2 \beta} - 3)-\sinh{2 \beta}\right]\,\gamma + O(\gamma^2),
\end{equation*}
and so does the symmetric logarithmic derivative
\begin{equation*}
    \logd = -\Delta\hams + l_1\,\op{\sigma}_{z,1} + O(\gamma^2),
\end{equation*}
with $l_1 = 1 + \tanh{\beta} + \sech^2{\beta}\,(\beta + 2 \beta\,\tanh{\beta} - 1)$. We know from Sec.~\ref{sec:optimality-general} that the classical Fisher information for measurements in the local energy basis and the QFI must agree in the zeroth and first order in $\gamma$. However, as noted in Sec.~\ref{sec:framework}, due to the coupling in this toy model being 'non-dissipative', i.e., $[\op{H}_\text{int},\op{H}] = 0$, $\logd$ is diagonal in the $\hams$ basis at all orders. Hence, energy measurements are always optimal in this particular case.

\subsection{\blue{Dissipative coupling}}
\label{app:counterexamples_B}

Let us now turn to a more general scenario. Namely,
\begin{equation*}
\op{H} = \hams + \hamb + \op{H}_\text{int} = \omega\,\op{\sigma}_{z,1}\otimes\id_2 + \omega\,\id_1\otimes\op{\sigma}_{z,2} + \gamma\,\op{\Pi}_{1} \otimes \op{\sigma}_{x,2},   \end{equation*}
where $\op{\Pi} = \ket{1}\bra{1}$. In this case, we can see that 
\begin{align*}
   \pist = \pis + \gamma\,p_1\,\op{\sigma}_{x,1} + \gamma^2\,p_2\,\op{\sigma}_{z,1} + O(\gamma^3), 
\end{align*}
with coefficients
\begin{align*}
    p_1 &= \frac{1}{4}\,\tanh{\beta}\,(\tanh{\beta} -1) \\
    p_2 &=\frac{1}{8}\,\sech{\beta}\,(\tanh{\beta}-1)\,(\beta\,\sech{\beta}-\sinh{\beta}).
\end{align*}
In particular, note how the mean-force Gibbs state does feature $\beta$-dependent coherences. 

The corresponding symmetric logarithmic derivative is
\begin{equation*}
    \logd = -\Delta\hams + \gamma\,l_1\,\op{\sigma}_{x,1} + O(\gamma^2),
\end{equation*}
with $ l_1 = \frac{1}{2}\,\sech^2{\beta}\,(2 \tanh{\beta} -1)$, meaning that the optimal measurement is non-diagonal in the energy basis already to first order in the coupling strength. And yet, as we proved in full generality in Sec.~\ref{sec:optimality-general}, the classical Fisher information of local energy measurements and the quantum Fisher information must agree to \textit{second} leading order. Indeed, direct calculation shows that
\begin{equation*}
    \fish-\cfish = \frac{1}{4}\,\sech^4{\beta}\,(1-2 \tanh{\beta})^2\,\gamma^2 + O(\gamma^3).
\end{equation*}

\renewcommand{\thefootnote}{\fnsymbol{footnote}}

\bibliographystyle{sirun}

\bibliography{references}

\end{document}